\documentclass[aps,prl,reprint,groupedaddress,showpacs,longbibliography]{revtex4-1}

\usepackage{siunitx}
\usepackage{amssymb,amsmath} 
\usepackage{graphicx}
\usepackage{xcolor}
\usepackage{pgfplots}
\usepgfplotslibrary{groupplots}

\usepackage{pgfplots}

\usepackage{stackrel}
\usepackage[normalem]{ulem}

\begin{document}

\title{Observation of vibrational dynamics of orientated Rydberg-atom-ion molecules}

\author{ Yi-Quan Zou$^{1,*}$, Moritz Berngruber$^{1,*}$, Viraatt S. V. Anasuri$^1$, Nicolas Zuber$^1$, Florian Meinert$^1$, Robert L\"ow$^1$, Tilman Pfau$^{1, \dag}$}

\affiliation{5. Physikalisches Institut and Center for Integrated Quantum Science and Technology, Universit\"at Stuttgart, Pfaffenwaldring 57, 70569 Stuttgart, Germany}

\date{\today}

\begin{abstract}
Vibrational dynamics in conventional molecules usually takes place on a timescale of picoseconds or shorter. A striking exception are ultralong-range Rydberg molecules, for which dynamics is dramatically slowed down as a consequence of the huge bond length of up to several micrometers. Here, we report on the direct observation of vibrational dynamics of a recently observed Rydberg-atom-ion molecule. By applying a weak external electric field of a few mV/cm, we are able to control the orientation of the photoassociated ultralong-range Rydberg molecules and induce vibrational dynamics by quenching the electric field. A high resolution ion microscope allows us to detect the molecule's orientation and its temporal vibrational dynamics in real space. Our study opens the door to the control of molecular dynamics in Rydberg molecules. 
\end{abstract}

\maketitle

Probing molecular dynamics with a high temporal and spatial resolution is of great interest in molecular physics and chemistry. Vibrational dynamics in regular molecules which occurs on ultrafast timescales has been observed and studied by making use of femtosecond lasers \cite{mokhtari1990direct, zewail2000femtochemistry, krausz2009attosecond}. 
The experimental advances in cooling atoms and molecules to ultralow temperatures have enabled the study of molecules and their dynamics at vastly different energy scales. Examples include the production of new types of molecules such as weakly bound Feshbach molecules \cite{kohler2006production} and Efimov states \cite{greene2017universal}, and the quantum control of molecular internal states \cite{carr2009cold, ospelkaus2010controlling}, dynamics \cite{kunitski2021ultrafast}, and cold chemical reactions  \cite{krems2008cold, balakrishnan2016perspective}.  Another candidate that shows unique properties for studying molecular dynamics is ultralong-range Rydberg molecules \cite{shaffer2018ultracold, fey2020ultralong}. These molecules exhibit extreme properties such as huge bond lengths and large electric dipole moments up to thousands of Debye \cite{li2011homonuclear, booth2015production}. Owing to their macroscopic size, Rydberg molecules allow us to probe nuclear dynamics in ``slow-motion" on the microsecond timescale which is accessible with standard laser and electronic techniques. In contrast to ordinary molecules which require strong electric fields for alignment \cite{stapelfeldt2003colloquium}, Rydberg molecules can also be aligned and orientated with relatively weak external fields.
Different types of Rydberg molecules have been experimentally observed, like Rydberg ground-state atom molecules \cite{greene2000creation, bendkowsky2009observation, bendkowsky2010rydberg, anderson2014photoassociation, booth2015production, niederprum2016observation, peper2021heteronuclear}, and Rydberg macrodimers consisting of two Rydberg atoms \cite{boisseau2002macrodimers, overstreet2009observation, samboy2011formation, sassmannshausen2016observation, hollerith2019quantum}. Their molecular properties, including photoassociation mechanisms \cite{kleinbach2017photoassociation, kiffner2012dipole} and interactions with external magnetic \cite{lesanovsky2006ultra, krupp2014alignment, hummel2018spin} and electric fields \cite{hummel2021electric}, have been studied. However, the direct observation of nuclear dynamics remained elusive.

\begin{figure}[t!]
\centering
\includegraphics[width=0.47\textwidth]{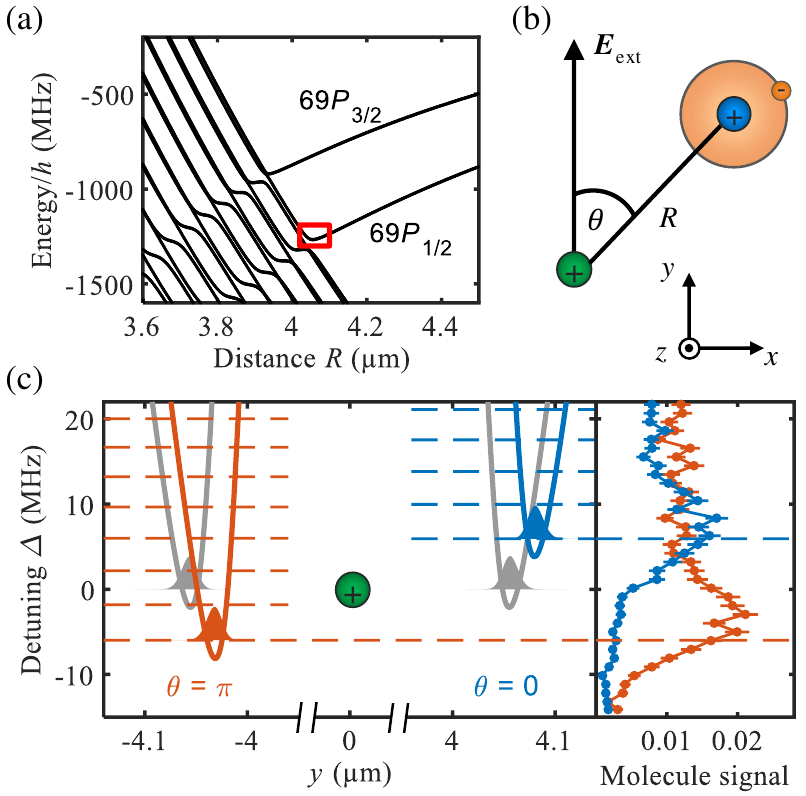}
\caption{(a) Potential energy curves of $^{87}$Rb${_2^{\,\,+}}$ as a function of internuclear distance $R$. $h$ is Planck's constant. Molecular potential wells are formed at the avoided crossings between the $69P$ states and the hydrogenic manifolds. The potential well marked with a red rectangle is the relevant binding potential studied in this paper.
(b) Rydberg-atom-ion molecule in an external electric field. $R$ is the internuclear distance and  $\theta$ the angle between the molecular axis and the external field ${\boldsymbol E}_{\rm ext}$.  In our experiment, ${\boldsymbol E}_{\rm ext}$ is pointing along the +$y$ direction. The imaging axis is along the $z$-axis. 
(c) Left panel: molecular potential along the $y$-axis in the presence of an external field of 10\,mV/cm (blue curve for $\theta = 0$ and orange curve for $\theta = \pi$). Gray curves represent the field free potential. Zero detuning refers to the lowest vibrational energy at zero external field. Shaded areas indicate the ground vibrational wavefunctions. Horizontal dashed lines show computed vibrational levels. Right panel: photoassociation spectroscopy for molecules orientated to  $\theta \approx 0 $ (blue) and $\theta \approx \pi $ (orange). Molecules detected within an angle window of $\pm0.2\pi$ are taken into account. Data points represent a bin of the raw data in 1\,MHz steps. Error bars denote the standard error of the mean.}
\label{fig1}%
\end{figure}

Most recently, a new type of Rydberg molecule consisting of a Rydberg atom and an ion has been proposed \cite{deiss2021long, duspayev2021long} and also experimentally observed in ultracold rubidium gases \cite{zuber2022observation}. The binding mechanism is based on the interaction between the ionic charge and the induced flipping dipole of the Rydberg atom in the field of the ion. Molecular bound states can be formed in the potential wells located at the avoided crossings between the $nP$ Rydberg state and the hydrogenic manifolds for atomic species with a noninteger quantum defect (see Fig.\,\ref{fig1}a). In Ref.\,\cite{zuber2022observation}, vibrational levels were identified by photoassociation spectroscopy, and the molecular bond length and alignment have been spatially resolved by an ion microscope \cite{veit2021pulsed}.

In this letter, we report on the direct experimental observation of vibrational dynamics in single Rydberg-atom-ion molecules. 
We make use of an external electric field to manipulate the molecular potential between the Rydberg atom and the ion. 
The molecules are photoassociated in an orientated manner by applying a constant electric field on the order of 10\,mV/cm. Subsequently, vibrational wavepacket dynamics in the molecular potential is triggered by a quench of the external field to zero. By using an ion microscope as a detection tool, molecular dynamics can be directly observed as a function of evolution time. The measured results are found to be in good agreement with classical as well as quantum mechanical simulations.

First, we discuss the mechanism behind the electric-field-induced molecular orientation and vibrational dynamics. The presence of external fields breaks the spherical symmetry of the molecular potential, leading to an adiabatic potential energy surface which depends on the internuclear distance $R$ and the relative angle $\theta$ between the molecular axis and the external field ${\boldsymbol E}_{\rm ext}$ (see Fig.\,\ref{fig1}b). For an external field of 10\,mV/cm pointing along $+y$-axis, the relevant cut through the molecule potential surface of the $69P_{1/2}$ Rydberg-atom-ion dimer along the external field axis (corresponding to $\theta = 0$ and $\theta = \pi$) is shown in the left panel of Fig.\,\ref{fig1}(c).  The potentials are calculated with a modified version of the pair interaction program introduced in Ref.\,\cite{weber2017calculation}.  
In general, the electric field introduces a displacement of the potential minima along the internuclear axis compared to the field-free potentials (gray curves). At the same time, the depth of the potential is shifted by about 6\,MHz. This shift originates from the higher order multipolar interactions (quadrupole, octupole, etc.) associated with the inhomogeneous field of the ion, and is proportional to the external field up to a few hundred mV/cm. The ratio between the energy shift and the external field gives an effective dipole moment of $\approx1.2$ kilo-Debye.

The experimental procedures to create and image single Rydberg-atom-ion molecules have been described in detail in Ref.\,\cite{zuber2022observation} and are briefly summarized here. The experiment starts with an ultracold cloud ($\SI{20}{\micro K}$) of about $10^6$ unpolarized $^{87}$Rb atoms in the ground state $|5S_{1/2}, F = 2\rangle$, with a peak density of $5\times10^{12}$\,cm$^{-3}$, trapped in a crossed dipole trap. The generation of single molecules is accomplished in two steps, first by two-photon ionization to create a single rubidium ion, and then by exciting a ground-state atom to the Rydberg state to photoassociate the molecule. For the photoassociation, two lasers at 780\,nm and 480\,nm which are 200\,MHz detuned from the intermediate $5P_{3/2}$ state are used. In this work, both lasers are linearly polarized along the $x$-axis such that the excitation probability is maximized for molecules aligned along the $y$-axis \cite{zuber2022observation}. Before detecting the molecule, we spatially separate the Rydberg atom and the ion along the optical axis ($z$-axis) by an electric field pulse which dissociates the molecule and accelerates the ion but does not ionize the Rydberg atom. In this way, the Rydberg atom and the ion can be distinguished by their arrival time on the detector. Finally, we field ionize the Rydberg atom and project both the Rydberg core and the ion onto our detector with a spatial resolution of 200\,nm. The detection of a Rydberg atom and an ion within one measurement cycle comprises a molecular event. 
In a single atom cloud, up to 6080 cycles of creation and detection are performed. For the chosen molecular state which is connected to the $69P_{1/2}$ asymptote, the molecular binding strength is weak enough such that the field dissociation during the separation process plays no significant role for the imaging of the molecule \cite{zuber2022observation}.

Orientated molecules are created by photoassociation in the presence of a weak external field  ${\boldsymbol E}_{\rm ext}=10$\,mV/cm \footnote{For the applied external electric fields, we only show the digits before decimal point. The stability of the electric field is better than 100 microvolts per centimeter.}. Energy shifts for two different molecular orientations with $\theta \approx 0$ and $\theta \approx \pi$, as can be seen in the right panel of  Fig.\,\ref{fig1}(c), allow us to distinguish them spectroscopically. The two spectra show line shifts in opposite directions with respect to the zero detuning. Here, only the lowest vibrational peaks are clearly visible, while higher vibrational levels are hardly resolvable. This can be explained by an excitation linewidth which is comparable to the vibrational spacing. The large linewidth is mainly caused by motional transient time broadening effects and the short pulse duration (0.5\,$\si{\micro s}$) for photoassociation. The shifts of the lowest vibrational peaks are consistent with the value given by theoretical calculations (blue and orange dashed lines).

\begin{figure}[t!]
\centering
\includegraphics[width=0.48\textwidth]{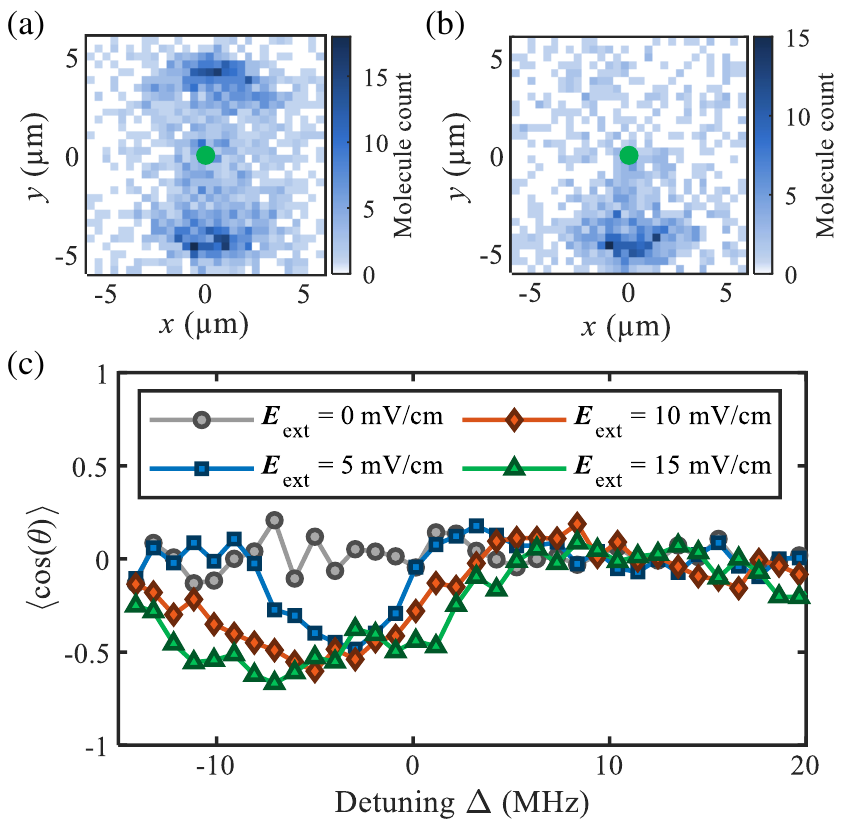}
\caption{ Molecular image and orientation. Image of molecules for ${\boldsymbol E}_{\rm ext}=0$\,mV/cm (a) and ${\boldsymbol E}_{\rm ext}=10$\,mV/cm (b). The corresponding detunings are 0\,MHz and $-6$\,MHz, which is resonant to the lowest vibrational level at $\theta = \pi$. $x$ and $y$ represent the relative position of the Rydberg atom with respect to the ion. (c) Molecular orientation $\bar O \equiv \langle \cos(\theta)\rangle$ as a function of the laser detuning for ${\boldsymbol E}_{\rm ext}=0, 5, 10$ and $15$\,mV/cm. Data points represent a bin of the raw data in 1\,MHz steps.}
\label{fig2}%
\end{figure}

Thanks to our spatially resolved detection, the molecular orientation can be directly identified by imaging the relative positions of the Rydberg atom and the ion, as shown in Figs.\,\ref{fig2}(a) and (b).
The images are shown in relative coordinates with the ion located at the origin. 
In the field free case (Fig.\,\ref{fig2}a), the photoassociated molecules are aligned along the $y$-axis due to the set polarization of the photoassociation lasers, but are not orientated. When the electric field is present, specific detunings of the laser allow the excitation of orientated molecules around $\theta = \pi$, as shown in Fig.\,\ref{fig2}(b).
For a more quantitative characterization, we define the molecular orientation as $\bar O \equiv \langle\cos(\theta)\rangle$ where $\langle \, \rangle$ denotes the expectation value in the imaging plane, and plot it as a function of the laser detuning $\Delta$ for ${\boldsymbol E}_{\rm ext}=0, 5, 10$ and $15$\,mV/cm in Fig.\,\ref{fig2}(c). The absolute magnitude of $\bar O$ represents the degree of orientation whereas its sign determines the direction of molecular orientation with respect to the external field. At zero field, no orientation is expected and $\bar O$ is found consistent with zero. In the presence of an external field, $\bar O$ varies with the laser detuning and features a negative minimum at a detuning $\approx -6$\,MHz where only the lowest vibrational state around $\theta = \pi$ is excited. The maximum degree of orientation increases with the strength of the external field. When the detuning is increased to be positive, the lowest vibrational state at $\theta = 0$ as well as higher vibrational states at $\theta = \pi$ can be excited, leading to a small positive value of $\bar O$. For very large positive and negative detunings, the orientation tends to be zero due to the homogeneous background signal. In contrast to techniques used for conventional molecules where strong fields are applied to orient molecules after the creation \cite{stapelfeldt2003colloquium}, here the ion-Rydberg molecules are directly excited in an oriented way by using a weak electric field. The detected orientation is limited by the excitation linewidth of the photoassociation process.

\begin{figure}[t!]
\centering
\includegraphics[width=0.5\textwidth]{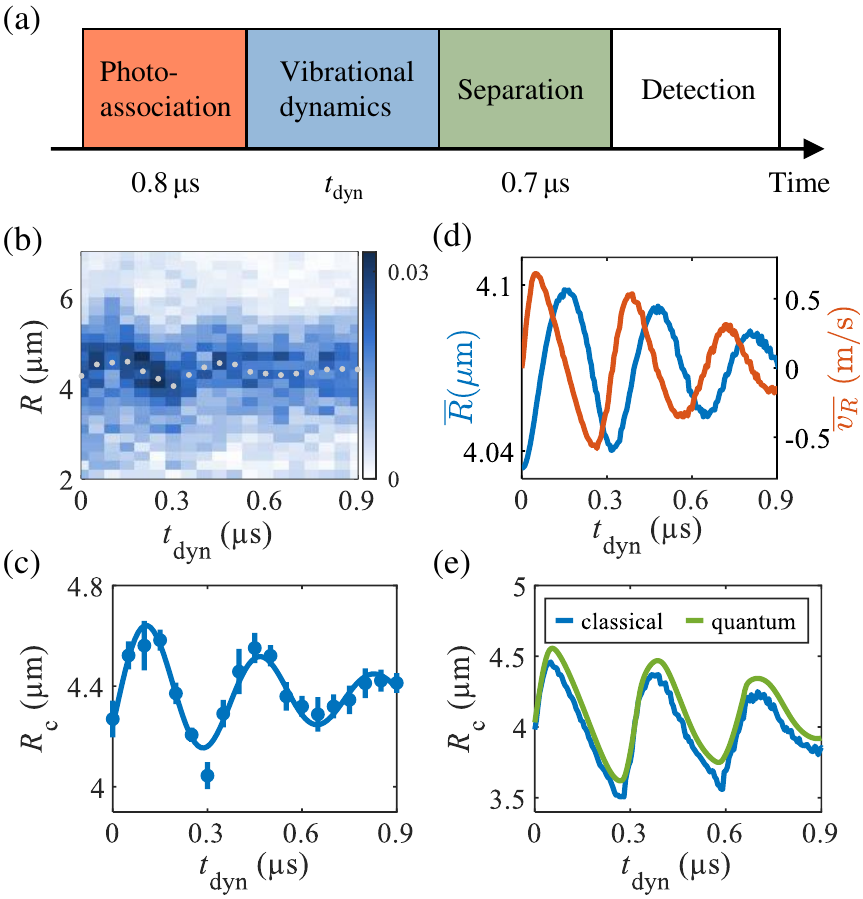}
\caption{Vibrational wavepacket dynamics. (a) Experimental sequence. (b) Radial density distribution of the wavepacket, integrated over an azimuthal segment of $0.4\pi$ around $\theta = \pi$,  detected after separation for variable dynamics time $t_{\rm dyn}$. White dots represent the fitted center $R_{\rm c}$ of the wavepacket with a Gaussian function. (c) Plot of the center $R_{\rm c}$. Error bars denote the fitting uncertainties. The solid line is a fit of the oscillation based on a damped sinusoid. (d) Calculated mean radial position $\overline{R}$ (blue line) and mean radial velocity $\overline{v_{\rm R}}$ (orange line) of the wavepacket before separation with a classical trajectory simulation. (e) Fitted center of the wavepacket after separation from theoretical simulations modeled by classical and quantum mechanical calculations. }
\label{fig3}%
\end{figure}

In a further step, we investigate the vibrational wavepacket dynamics within the molecular potential. The experimental procedure is sketched in Fig.\,\ref{fig3}(a). The molecules were initially prepared orientated to $\theta \approx \pi$ at ${\boldsymbol E}_{\rm ext}=10$\,mV/cm with $\Delta = -6$\,MHz. The vibrational dynamics is triggered by a sudden quench of the external field to zero, which leads to a radial displacement of the molecular potential (see Fig.\,\ref{fig1}c). The external field is switched off within 10\,ns. On this timescale, the nuclear motion is frozen whereas the electronic wavefunction follows adiabatically the variation of the electric field. Consequently, the initial vibrational wavepacket is no longer at the equilibrium position and starts to oscillate in the potential. After a variable evolution time $t_{\rm dyn}$, we separate and detect the nuclear constitutes of the molecule as previously described. The separation duration is set to be 0.7\,$\si{\micro s}$ in our experiment such that the Rydberg core and the ion are well distinguished by their arrival times on the detector.

Figure\,\ref{fig3}(b) shows the detected radial density distribution of the wavepacket, integrated over  an azimuthal segment of $0.4\pi$ around $\theta = \pi$, as a function of the evolution time. The center of the wavepacket $R_{\rm c}$ is extracted with a Gaussian fit and is plotted in Fig.\,\ref{fig3}(c). An oscillation with an amplitude of a few hundred nanometers is observed.
To model the vibrational dynamics, a classical trajectory simulation is performed to calculate the mean radial position $\overline{R}$ (blue curve in Fig.\,\ref{fig3}d) and mean radial velocity $\overline{v_{\rm R}}$ (orange curve) of the wavepacket before the separation (see supplementary materials \cite{REFSM}). The calculated oscillating amplitude of $\overline{R}$ is much smaller than the detected amplitude of $R_{\rm c}$.
In addition, the experimentally observed evolution of $R_{\rm c}$, especially the initial phase of the oscillation, matches better with the velocity curve than the position curve.
A full simulation of the molecular evolution before the detection reveals the influences of the finite separation time. 
The obtained wavepacket center after the separation, as shown by the blue curve in Fig.\,\ref{fig3}(e), exhibits an oscillation in phase with the calculated $\overline{v_{\rm R}}$ before separation. This can be explained by the fact that the wavepacket propagates freely in the object plane during the separation after the almost instant dissociation of the molecule. 
An additional one-dimensional quantum mechanical simulation is also performed to calculate the wavepacket center $R_{\rm c}$ after the separation, as shown by the green curve in Fig.\,\ref{fig3}(e) (supplementary material).  Both the classical and the quantum simulation resemble our experimental results. The observed damped oscillation is a mapping of the evolution of velocity during the vibrational dynamics.
The decay of the oscillation can be mostly attributed to the anharmonicity of the molecular potential. Quantum revivals are expected to occur at later times around 6.6\,$\si{\micro s}$, but their experimental observation is hindered by the short lifetime of the molecule which was measured to be $2.6\pm0.1$\,$\si{\micro s}$ and might be limited by radiofrequency-induced transition to unbounded Rydberg states \cite{zuber2022observation}. Fitting the experimental data with a damped sinusoidal model yields a frequency of $2.8\pm0.1$\,MHz (blue solid line in Fig.\,\ref{fig3}c), close to the value of 3.1\,MHz given by the simulations shown in Fig.\,\ref{fig3}(e). The observed amplitude is smaller than that obtained in our simulations. A possible explanation of this discrepancy is that the dissociation of the molecule is treated as an instant process in the simulation, while in real experiment it takes place on a timescale of 10\,ns.

\begin{figure}[t!]
\centering
\includegraphics[width=0.47\textwidth]{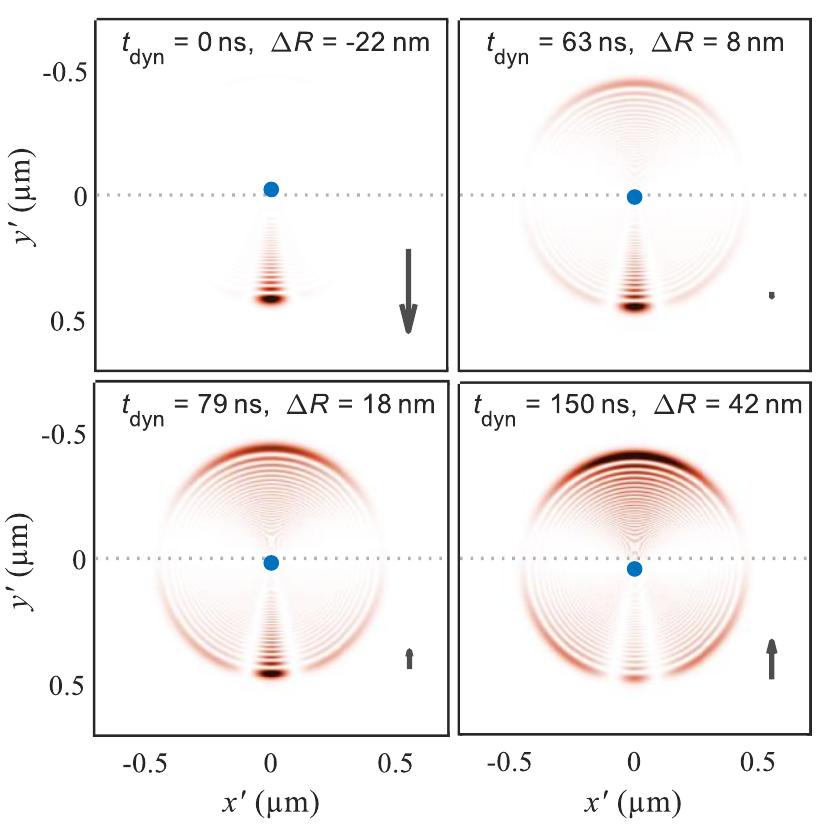}
\caption{Rydberg electron dynamics during the periods of nuclear vibration.  Dotted lines mark the equilibrium position of the Rydberg atom's core. Blue dots indicate the position of the Rydberg core at different evolution times $t_{\rm dyn}$. $\Delta R$ is the displacement of the Rydberg core relative to the equilibrium position. Arrows show the direction and relative magnitude of the Rydberg atom's dipole moment.  The ion is located on the $y$-axis about 4\,{$\si{\micro m}$} above the Rydberg core, which is not shown in the figure.}
\label{fig4}%
\end{figure}

Finally, we look at the evolution of the Rydberg electron during the vibrational dynamics. Figure\,\ref{fig4} shows four snapshots of the calculated Rydberg electronic density distribution in the first half period of the dynamics. The electron vibrates with a much larger amplitude than the nuclei and the electronic density profiles vary dramatically as the nuclei oscillate around the equilibrium position $R_0$. When the Rydberg core is at the inner turning point with $\Delta R \equiv \overline{R} - R_0$ being negative (top left figure), the Rydberg electron is strongly localized in the azimuthal direction because of the high angular momentum character of the electronic wavefunction.  In this case, the induced dipole of the Rydberg atom points away from the ion. At the outer turning point (bottom right figure), the dipole moment points towards the ion with a broader angle distribution for the electron due to the dominant low angular momentum ($S$, $P$ and $D$) states.

In conclusion, we have studied the vibrational dynamics in single Rydberg-atom-ion molecules. We showed that molecular orientation and dynamics can be controlled via a weak electric field. We observed oscillations of the wavepacket within the molecular potential with a frequency orders of magnitude slower than that observed in conventional molecules. Our study can be extended in the following aspects. First, molecular states with a lower Rydberg principal quantum number can be studied with an optimized dissociation pulse with which the spatial imaging is not distorted. There, the vibrational dynamics is faster and less limited to the rather short lifetime of the molecule \cite{duspayev2022nonadiabatic}. As a result, quantum revivals should be observable. Second, in this work the dynamics is studied by manipulating the molecular potential with an electric field. An alternative way to induce molecular dynamics is to use microwaves which allow for excitation of various non-equilibrium states.
Finally, our result opens the path to experimental investigation of more sophisticated dynamics in Rydberg molecules, such as beyond Born-Oppenheimer dynamics in the vicinity of conical intersections \cite{hummel2021synthetic}. Molecular dynamics in Rydberg trimers enables the study of interesting three-body interactions and leads to a completely new regime of dynamics. Even more, with Rydberg molecules, a real-time tracing of chemical reaction processes on ultraslow timescales is possible \cite{schlagmuller2016ultracold}.

\begin{acknowledgments}
We thank O. A. Herrera-Sancho for discussions and proof reading of the manuscript. This work was supported by the Deutsche Forschungsgemeinschaft (DFG) [Project No. Pf 381/17-1 and No. Pf 381/17-2] as part of the SPP 1929 ``Giant Interactions in Rydberg Systems (GiRyd)" and has received funding from the Bundesministerium für Bildung und Forschung (BMBF) - project ``Theory-Blind Quantum Control - TheBlinQC" (grant no. 13N14847). The project received funding from the
European Research Council (ERC) under the European Union’s Horizon 2020 research and innovation programme (Grant agreement No. 101019739 - LongRangeFermi). F. M. acknowledges funding from the Federal Ministry of Education and Research (BMBF) under the grant CiRQus.
\\
\end{acknowledgments}

*These authors contributed equally to this work.

$^\dag$ t.pfau@physik.uni-stuttgart.de

\bibliography{bib_MolecularDynamics}

\pagebreak
\clearpage

\begin{center}
\textbf{\large Supplemental Material}
\end{center}

\subsection{Classical trajectory simulation}

In this section, we describe the details of classical trajectory simulation performed to obtain the results shown in Fig.\,3(d) and (e) in the main text. The relative motion between the Rydberg atom and the ion of the molecule is numerically simulated for processes from photoassociation to separation (see Fig.\,3a). For the detection, charged particle trajectory simulations \cite{veit2021pulsed} of the ion microscope have shown a high spatial resolution detection with a field depth of more than 1\,mm. 

The initial relative position of Rydberg atom with respect to the ion is randomly chosen from a spatially dependent excitation probability. Here we restrict the molecular orientation in the range of $\theta = \pi\pm 0.2\pi$. Within this angle, we assume that the angle dependent excitation probability is the same as that of the field free case which is determined by the polarization of the photoassociation lasers \cite{zuber2022observation}. The radial distribution is calculated with the molecular potential of $\theta = \pi$ with  ${\boldsymbol E}_{\rm ext} = 10$\,mV/cm, taking into account the excitation linewidth of the photoassociation and the mixed fraction of $S$ and $D$ states.
The initial velocities of the Rydberg atom and ion are randomly sampled over a Maxwell-Boltzmann distribution for a temperature of $\SI{20}{\micro K}$. 

The simulation of the relative motion starts immediately after the excitation of the Rydberg atom at a time uniformly sampled over the 0.8\,$\si{\micro s}$ long photoassociation process. Since the molecule possesses a charge at one end of the bond, the external field can cause rotational motion for the whole molecule which is also considered in our simulation. When the electric field is quenched to trigger the vibrational dynamics, the interaction between the Rydberg atom and ion is switched to the field free molecular potential. The mean relative radial distance $\overline{R}$ and mean relative radial velocity $\overline{v_R}$ at the end of the vibrational dynamics are calculated and plotted as a function of $t_{\rm dyn}$ in Fig.\,3(d). For the separation process, we have considered the finite rising time of the dissociation electric field pulse which is faster than 10\,ns. The interaction is switched off when the external field reaches the same value as the field of the ion at the binding length. The positions of the Rydberg atom and ion at the end of the separation process are then projected onto the object plane, as happened in our experiment.  With the final two dimensional (2D) distribution of the relative position, we can obtain the radial profile by integrating over the azimuthal direction and extract the radial center $R_{\rm c}$ with a Gaussian fit. We plot $R_{\rm c}$ as a function of $t_{\rm dyn}$ in Fig.\,3(e) (blue curve). \\

\subsection{Quantum wavepacket simulation}

In addition to the classical trajectory simulation, a simple one dimensional (1D) quantum mechanical calculation is also performed to simulate the evolution of the wavepacket before the final detection. In this case, we assume the initial wavepacket is prepared in the lowest vibrational state of the molecular potential of $\theta = \pi$ with ${\boldsymbol E}_{\rm ext} = 10$\,mV/cm. The initial wavepacket first evolves in the field free molecular potential during the vibrational dynamics, and then propagates and expands freely during the separation process. The density distribution at the end of the separation is fitted with a Gaussian function to extract the center of the wavepacket $R_{\rm c}$ which is then plot as a function of $t_{\rm dyn}$ in Fig.\,3(e) (green curve). The result of the 1D quantum wavepacket simulation agrees well with that of the classical trajectory simulation, indicating that the rotational motion of the molecule, 2D projection of the image and the dissociation process play no significant role in the detection of vibrational dynamics.

\end{document}